\documentclass[aps,prl,reprint,superscriptaddress,floatfix]{revtex4-2}

\usepackage{graphicx}
\usepackage{amsmath,amssymb,amsfonts,bm}
\usepackage{amsthm}
\usepackage{mathrsfs}
\usepackage{xcolor}
\usepackage{booktabs}
\usepackage{siunitx}
\usepackage{dcolumn}
\usepackage{multirow}
\usepackage{braket}
\usepackage[normalem]{ulem}
\usepackage{textcomp}
\usepackage{listings}
\usepackage{soul}
\usepackage{comment}
\usepackage{algpseudocode}
\usepackage[normalem]{ulem}
\usepackage[colorlinks=true, linkcolor=black, citecolor=black, urlcolor=blue]{hyperref}

\renewcommand{\include}[1]{\input{#1}}

\newcommand{\unizar}{Centro de Astropart\'{\i}culas y F\'{\i}sica de Altas Energ\'{\i}as (CAPA), Universidad de Zaragoza, Pedro Cerbuna 12, 50009 Zaragoza, Spain}
\newcommand{\LSC}{Laboratorio Subterr\'aneo de Canfranc, Paseo de los Ayerbe s.n., 22880 Canfranc Estaci\'on, Huesca, Spain}

\begin{document}

\title{First Experimental Bounds on Transverse Plasmon Solar Axions with ANAIS-112}

\author{S.~J.~Hollick}
\thanks{Corresponding author.}
\email{shollick@unizar.es}
\affiliation{\unizar}
\affiliation{\LSC}

\author{M.~Giannotti}
\thanks{Corresponding author.}
\email{mgiannoti@unizar.es}
\affiliation{\unizar}

\author{J.~Ruz}
\thanks{Corresponding author.}
\email{Jaime.Ruz@tu-dortmund.de}
\affiliation{
Fakult{\"a}t f{\"u}r Physik,~
Technische~Universit{\"a}t~Dortmund,
44221~Dortmund,~Germany}
\affiliation{\unizar}
\affiliation{
Cluster of Excellence “Color meets Flavor”, Technische Universit{\"a}t Dortmund}

\author{J.~K.~Vogel}
%\email{Julia.Vogel@tu-dortmund.de}
\affiliation{
Fakult{\"a}t f{\"u}r Physik,~
Technische~Universit{\"a}t~Dortmund,
44221~Dortmund,~Germany}
\affiliation{\unizar}
\affiliation{
Cluster of Excellence “Color meets Flavor”, Technische Universit{\"a}t Dortmund}

\author{R.~M.~Alkaddah}
%\email{rama.alkaddah@tu-dortmund.de}
\affiliation{
Fakult{\"a}t f{\"u}r Physik,~
Technische~Universit{\"a}t~Dortmund,
44221~Dortmund,~Germany}

\author{J.~Amar{\'e}}
\affiliation{\unizar}
\affiliation{\LSC}
 \author{J.~Apilluelo}
 \affiliation{\unizar}
\affiliation{\LSC}
\author{S.~Bharat}
\affiliation{\unizar}
\affiliation{\LSC}
 \author{S.~Cebri{\'an}}
 \affiliation{\unizar}
\affiliation{\LSC}
\author{D.~Cintas}
\affiliation{\unizar}
\affiliation{\LSC}
 \author{I.~Coarasa}
 \affiliation{\unizar}
\affiliation{\LSC}
\author{E.~Garc\'{\i}a}
\affiliation{\unizar}
\affiliation{\LSC}
%\author{S.~J.~Hollick}
%\affiliation{\unizar}
%\affiliation{\LSC}
\author{M.~Mart\'{\i}nez}
\affiliation{\unizar}
\affiliation{\LSC}
\author{Y.~Ortigoza}
\affiliation{\unizar}
\affiliation{\LSC}
\affiliation{Escuela Universitaria Polit\'ecnica de La Almunia de Do\~na Godina (EUPLA), Universidad de Zaragoza, Calle Mayor 5, La Almunia de Do\~na Godina, 50100 Zaragoza, Spain}
\author{A.~Ortiz~de~Sol{\'o}rzano}
\affiliation{\unizar}
\affiliation{\LSC}
\author{T.~Pardo}
\affiliation{\unizar}
\affiliation{\LSC}
\author{J.~Puimed{\'o}n}
\affiliation{\unizar}
\affiliation{\LSC}
\author{M.~L.~Sarsa}
\affiliation{\unizar}
\affiliation{\LSC}
\author{C.~Seoane}
\affiliation{\unizar}
\affiliation{\LSC}

%\collaboration{ANAIS-112 Collaboration}

%%%%%%%%%%%%%%%

\date{\today}

\begin{abstract}
Solar axion-like particles (ALPs) can be resonantly produced through the conversion of
transverse plasmons in the magnetic
field of the solar interior, introducing a production mechanism complementary to the
conventional Primakoff process.
%\SH{Susana: This could be defined also in abstract}
%\mg{I would rather drop BCA from the abstract than define it here. BCA constrains $g_{ae}$, whereas this Letter is entirely a $g_{a\gamma}$ analysis: TP production, Primakoff production and inverse-Primakoff detection. Mentioning it among the ``conventional'' processes in the abstract is slightly misleading, and it costs us a clause we need for the result. The BCA discussion belongs in the introduction, where the COSINE-100 comparison motivates it.}
In this Letter, we report the first experimental search for solar ALPs produced through this mechanism. Using an exposure of 625.75\,kg$\times$yr from ANAIS-112, we search for the annual modulation of the flux induced by the variation of the Earth--Sun distance, assuming detection
through inverse Primakoff conversion in the detector.
Modulation amplitudes fitted in 1\,keV bins between 1--20\,keV are consistent with the absence of modulation, yielding the first constraints on the axion--photon coupling
that include the resonant transverse-plasmon flux,
$g_{a\gamma} < 1.32\times 10^{-9}\,\mathrm{GeV}^{-1}$ (90\%~C.L.) at $m_a \simeq 150$\,eV.
%\mg{I think the limit and the mass at which it is reached must appear in the abstract. It is not competitive with HB stars or CAST, but what we claim as new is the channel, not the strength.} 
This previously unexplored production channel extends the sensitivity of ANAIS-112 in the mass range 35--280\,eV, where, for the expected solar magnetic-field models, the transverse-plasmon flux exceeds the Primakoff flux by up to two orders of magnitude. %\mg{I refined some terminology. Eliminated ``demonstrate'' a mass region. The region is a prediction of the flux calculation. Also, I think ``contribute more than'' understates Figs.~2 and C, which show up to two orders of magnitude, so I prefer to give the number. }.

\end{abstract}

\keywords{axions, dark matter, ALPs, annual modulation, plasmon ...}
%Use showkeys class option if keyword                
%display desired
\maketitle
\textit{Introduction.---}\label{sec:intro}
Axions and axion-like particles (ALPs) remain among the best-motivated extensions of the Standard Model, with laboratory, astrophysical, and cosmological searches probing their couplings to photons, electrons, and nucleons~\cite{PQ,Weinberg,Wilczek,Jaeckel2010,Marsh2016,Irastorza2018}. Among these, solar axions provide one of the most sensitive probes of the axion--photon interaction through their production in the solar interior and subsequent detection in terrestrial experiments~\cite{Sikivie:1983ip}.

Solar axion searches have traditionally focused on production through the Primakoff process~\cite{Raffelt1986,Raffelt1990} and, more recently, on electron-induced mechanisms involving Compton scattering, bremsstrahlung, atomic recombination, and atomic de-excitation (BCA)~\cite{Redondo:2013wwa}. 
It has since been shown that solar magnetic fields can resonantly convert transverse plasmons into axions whenever the plasma frequency matches the axion mass, $\omega_p(r)=m_a$~\cite{Guarini2020}.
%\mg{Two ``More recently'' in consecutive sentences.}
This mechanism produces a qualitatively new solar axion flux whose normalization and spectral shape are determined by the magnetic structure and plasma properties of the solar interior. Despite its potentially large enhancement over the conventional Primakoff flux, this production channel has not previously been explored experimentally.

In this Letter we report the first experimental search for solar axions produced through resonant transverse-plasmon conversion using the annual modulation expected from the variation of the Earth--Sun distance in the ANAIS-112 experiment. 
Annual modulation searches for solar axions were first performed by the COSINE-100 collaboration using the BCA produced solar axions as detected through the axio-electric channel, placing limits on $g_{ae}$, the axion-electron coupling~\cite{COSINEannualSolar}. 
In this work, we extend this strategy to the axion--photon coupling $g_{a\gamma}$, by detection through the inverse Primakoff channel, utilizing the Primakoff flux in conjunction with the recently proposed transverse-plasmon (TP) solar axion flux. 

No statistically significant modulation is observed. We therefore derive the first experimental constraints on the axion--photon coupling associated with resonant TP production. %As shown in Fig.~\ref{fig:exclusion}, our limits improve upon the previous best laboratory constraints in the axion mass range 80--220~eV by up to 22\%, corresponding to a factor of 1.28 in the excluded coupling.
Figure~\ref{fig:exclusion} illustrates the axion-mass range, 35–280 eV, over which the TP flux exceeds the Primakoff flux by up to two orders of magnitude, resulting in an improvement of up to a factor of 2.73 in the excluded axion–photon coupling, corresponding to a $63.4\%$ lower upper limit on $g_{a\gamma}$.
%\mls{lower upper limit sound strange. moreover is redundant giving 2.73 factor of improvement and 63.4\%}

\begin{figure}[!h]
    \centering
    \vspace{-1 cm}
    \includegraphics[width=1.0\columnwidth]{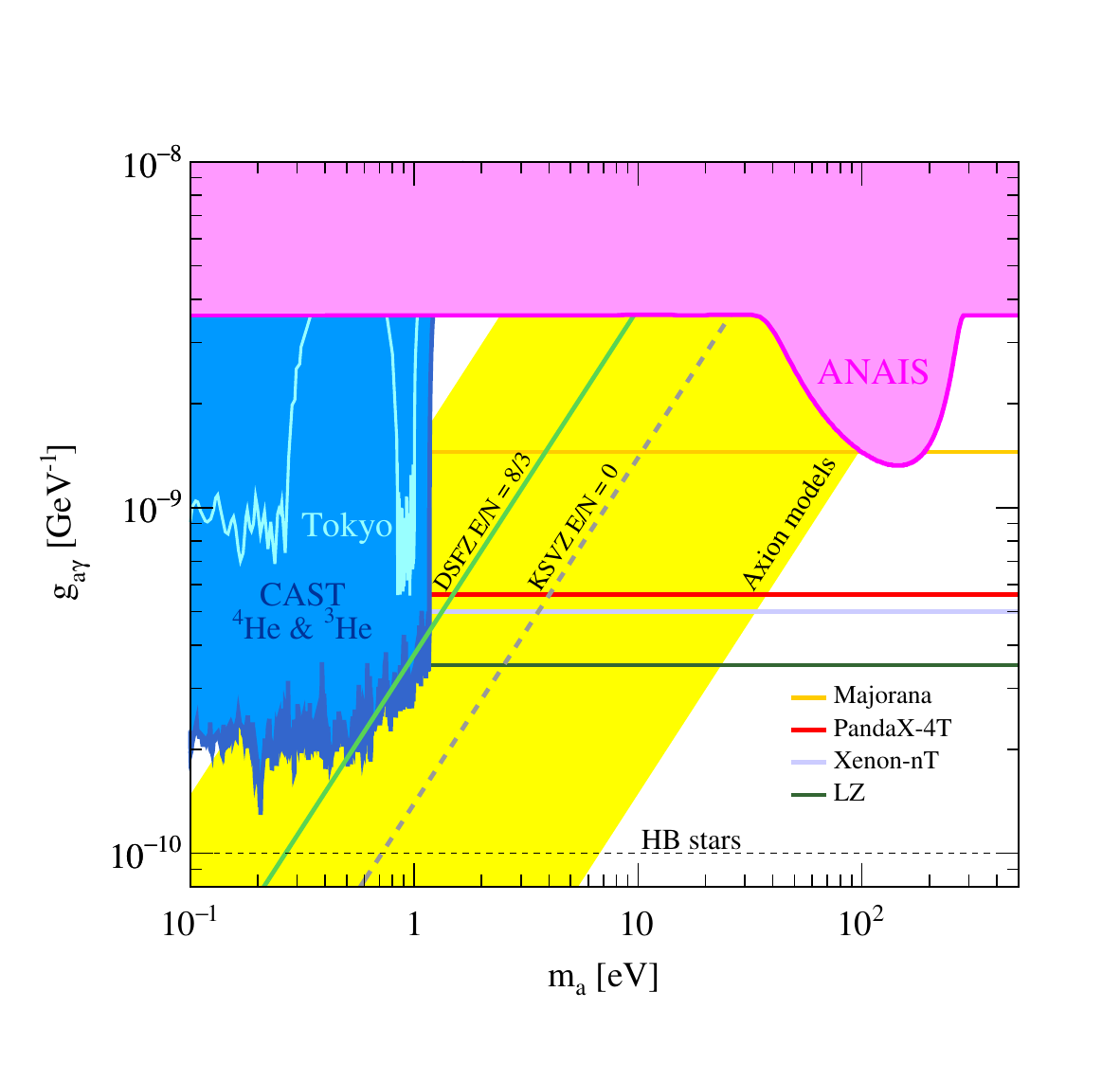}
    \vspace{-1. cm}
    \caption{
Exclusion limits at the 90\% C.L. on the axion--photon coupling $g_{a\gamma}$ as a function of the axion mass obtained with the 625.75~kg$\times$yr exposure of ANAIS-112 using the inverse Primakoff
detection channel (magenta region). The enhanced sensitivity in the 35--280~eV mass range arises from resonant transverse-plasmon axion production in the solar magnetic field, while outside this interval the
limits are determined by the standard Primakoff solar flux. The yellow band denotes benchmark QCD axion models. Existing limits from the CAST helioscope~\cite{CAST,CAST2024}, Majorana~\cite{Majorana}, and liquid-xenon TPC experiments PandaX-4T~\cite{PandaX_2025}, XENONnT~\cite{XENONnT_2022}, and LZ~\cite{LZ_2025} are shown for comparison.
%\mg{The caption as written attributes the whole curve to transverse-plasmon production, but outside 35--280~eV the limit is entirely Primakoff-driven.
%\mls{Majorana limit is shown, it should be provided the corresponding reference in the caption as those from LXe experiments}}
}
    \label{fig:exclusion}
\end{figure}
%
%%%%%%%%%%%%%%%%%%%
%
%%%%%%%%%%%%%%%
%%%%%%%%%%%%%%%%%%%%%%%%%%%%%%
%\vspace{0.2cm}
%\textit{Transverse-plasmon axions from the Sun.---}\label{sec:TP}
\textit{Signal prediction.---}\label{sec:TP}
The expected solar axion signal is computed by combining the conventional Primakoff production mechanism with the resonant conversion of transverse plasmons in solar magnetic fields~\cite{Guarini2020}. 
The Primakoff flux is evaluated with the \texttt{SolarAxionFlux} package~\cite{Hoof} adopting the B16-AGSS09 solar model~\cite{Bahcall_2006}, and the same model is used for the electron density and temperature profiles entering the transverse-plasmon calculation below.
%\mg{IMPORTANT: Please, verify what i wrote above. Which solar model did the runs actually use? We never say it, not even in the Supplemental Material, which only names the magnetic-field model. This is important as we need the plasma frequency (hence $n_e(r)$) profile.}
In a plasma, photons acquire an effective mass determined by the plasma frequency,
\begin{equation}
\omega_p=\sqrt{\frac{4\pi\alpha n_e}{m_e}},
\end{equation}
where $\alpha$ is the fine-structure constant, $n_e$ is the electron number density, and $m_e$ is the electron mass. Resonant plasmon--axion conversion occurs at the radius satisfying
\begin{equation}\label{eq:omegap_res}
\omega_p(r_{\rm res}) = m_a,
\end{equation}
where $m_a$ is the axion mass and $r_{\rm res}$ denotes the resonance radius. Consequently, each axion mass probes a distinct layer of the solar interior.

Including the finite transverse-plasmon width, the differential TP axion flux at Earth is given by~\cite{Guarini2020,Hoof}
\begin{equation}
\label{eq:TP_flux}
\begin{split}
\frac{d\Phi_a}{dE_a}
&=
\frac{R_\odot^3E_a^2}{2\pi^2D_\odot^2}
\int_0^1 dr\,r^2 \\
&\times
\frac{2\,\Gamma_T(E_a,r)\,
g_{a\gamma}^2|\mathbf{B}(r)|^2/3}
{\left[\Delta_p(E_{a},r)-\Delta_a(E_{a})\right]^2+\Gamma_T^2(E_a,r)/4} \\
&\times
\frac{1}{e^{E_a/T(r)}-1}.%\,
%W(r).
\end{split}
\end{equation}
%\mg{IMPORTANT: $W(r)$ should not be part of the flux equation. The flux is experimental independent. It does not depend on W. As it stands, Eq.~(3) is not the physical flux but our estimator of it. I would take it out of this equation and leave it just in Eq.~(S2); there we can describe it as a numerical prescription, together with a statement that the result is stable against variations of $W_0$.}
Here, $R_\odot$ is the solar radius, $D_\odot$ the mean Earth--Sun distance, and $r\equiv R/R_\odot$ the dimensionless radial coordinate within the Sun.
The prefactor accounts for the geometric dilution of the isotropically emitted axion flux between the solar surface and the Earth, while the integration over $r^2dr$ sums the contributions from all solar shells. The Bose--Einstein factor $[e^{E_{a}/T(r)}-1]^{-1}$ gives the occupation number of thermal transverse plasmons at the local temperature $T(r)$. The quantity $\Gamma_T(E_{a},r)$ denotes the damping rate of the transverse plasmon mode, proportional to the inverse photon mean free path~\cite{Redondo:2013lna}, which regulates the finite width of the resonance.
%\mg{Careful. Here we call this the damping rate $\Gamma_T$ while in the Supplemental Material we call the same quantity the absorption rate $\Gamma_{\rm abs}$ and define it as $\lambda_{\rm mfp}^{-1}$ (see Fig.~B).}. 
The resonant conversion probability is described by the resulting Breit-Wigner factor, where
\begin{equation}
\label{eq:deltas}
\Delta_p(E_a,r)
=
-\frac{\omega_{\rm p}^2(r)}{2E_a},
\qquad
\Delta_a(E_a)
=
-\frac{m_a^2}{2E_a},
\end{equation}
are the plasma and axion contributions to the oscillation wave number, respectively. Resonant enhancement occurs when $\Delta_p=\Delta_a$. %corresponding to the condition $\omega_{\rm p}(r)=m_a$, where the effective plasmon mass equals the axion mass \SH{Susana: Information in this sentence is already given by Eq. (2), right?}.
The production rate is proportional to the square of the axion--photon coupling, $g_{a\gamma}^2$, and to the local magnetic-field strength, with the factor $1/3$ arising from the angular average of the transverse magnetic-field component, $\langle B_\perp^2\rangle=|\mathbf{B}|^2/3$~\cite{Guarini2020}. 
%Finally, $W(r)$ denotes the numerical Gaussian window used to sample efficiently the narrow resonant region. It is centered on the resonance radius and chosen sufficiently broad compared with the physical resonance width so that it does not modify the integrated resonant axion flux, but only improves the numerical evaluation of the radial integral.\mg{As I mentioned above, perhpas this part should be moved to the appendix and here we should avoid mentioning W.}

Figure~\ref{fig:TP_fluxes_main} compares representative TP axion spectra with the Primakoff flux. Unlike the Primakoff spectrum, the TP flux depends strongly on the axion mass, with both its normalization and characteristic energy determined by the plasma and magnetic-field properties at the resonance radius. As the resonance condition is satisfied at different depths for different axion masses, resonant conversion moves from the solar tachocline, the thin shear layer separating the convective and radiative zones, to progressively deeper layers of the radiative interior. The shaded bands indicate the uncertainty associated with the adopted solar magnetic-field models ($\pm50\%$). Despite this uncertainty, the resonant TP flux exceeds the Primakoff flux by up to two orders of magnitude over a broad range of axion masses, substantially enhancing the sensitivity of laboratory searches.

\begin{figure}[!h]
    \centering
    \includegraphics[width=1.\columnwidth]{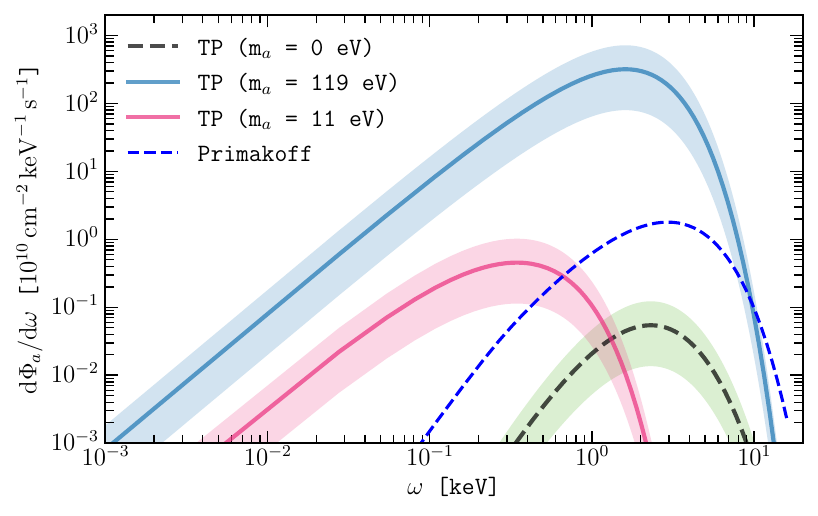}
    \caption{
    Differential solar axion fluxes for representative TP production scenarios compared with the conventional Primakoff flux (blue dashed line), assuming $g_{a\gamma}=5\times10^{-11}\,\mathrm{GeV}^{-1}$. The resonant TP spectra correspond to $m_a=11$~eV (tachocline, pink) and $m_a=119$~eV (radiative zone, blue), while the black curve shows the non-resonant ($m_a=0$) case. The shaded bands indicate the uncertainty associated with the adopted solar magnetic-field models.
    }
    \label{fig:TP_fluxes_main}
\end{figure}
%
%%%%%%%%%%%%%%%%%%%%%%%%%%%%%%%%%%%%%%%%%%%%%%%%%%%%%%%%%%%%%%%%%%%%%%%%%%
%\vspace{0.2cm}
\textit{Experiment and Search Strategy---}\label{sec:halo}
ANAIS-112 is a dark matter direct detection experiment located at Laboratorio Subterr\'aneo de Canfranc (LSC) in Spain, under a rock overburden of 2450 meter-water-equivalent. The experiment consists of nine \SI{12.5}{kg} NaI(Tl) scintillating crystals, providing a total active mass of \SI{112.5}{kg}, and has been operated from August 3, 2017, to January 15, 2026. The present analysis uses the 6.4-year data release, corresponding to an exposure of 625.75~kg$\times$yr~\cite{ANAIS6y}.

%Originally designed to test the annual modulation signal reported by DAMA/LIBRA, ANAIS-112 combines one of the lowest backgrounds achieved by a NaI(Tl)-based experiment with excellent long-term stability~\cite{ANAISPerformance,ANAIS6y}. 
%These characteristics make it particularly well suited to searches for solar axions through the annual modulation induced by the variation of the Earth--Sun distance. 
The detectors are surrounded by passive and active shielding, including archaeological and low-activity lead, an anti-radon enclosure, active muon vetoes, polyethylene shielding, and water tanks~\cite{ANAIS1yr}. Detector response has been continuously monitored through biweekly calibrations with external $^{109}$Cd and $^{252}$Cf sources and internal $^{40}$K and $^{22}$Na contamination, providing scintillation events within the ROI tagged by the coincidence with a high energy gamma.
This has allowed for accurate energy and efficiency calibrations
of the region of interest for the testing of the DAMA/LIBRA annual modulation \mbox{(1--6\,keV)}, as well as analysis of associated systematics~\cite{ANAIS6y}.

%The NaI(Tl) detectors are surrounded by passive and active shielding consisting of archaeological lead, low-activity lead, an anti-radon enclosure, active muon vetoes, polyethylene shielding, and water tanks~\cite{ANAIS1yr}. Detector performance has been regularly monitored through periodic calibrations with $^{109}$Cd and $^{252}$Cf sources~\cite{ANAIS6y}.

%NaI(Tl)-based dark matter experiments are particularly well suited to searches for annually modulated signals from solar axions. The first such search was performed by COSINE-100, which searched for an annual modulation induced by the conventional solar axion flux~\cite{COSINEannualSolar}. Building on this approach, the present work considers the transverse-plasmon solar axion flux and presents the first experimental search targeting this production mechanism, utilizing the inverse Primakoff conversion to directly probe the $g_{a\gamma}$ coupling. 

%\JRA{Maybe mentioning KSVZ ALPs for these mass ranges?}
%
%%%%%%%%%%%%%%%%%%%%%%%%%%%%%%%%%%%%%%%%%%%
%%%%%%%%%%%%%%%%%%%%%%%%%%%%%%%%%%%%%%%%%%%
%%%%%% Inverse Primakoff  Detection Channels %%%%%
%\vspace{0.2cm}
%\textit{Inverse Primakoff Detection Channel.---}
 
Solar axions coupled to photons can be detected in ANAIS-112 through the inverse Primakoff effect, in which an incident axion converts coherently into a photon in the Coulomb field of an atomic nucleus. Because the momentum transfer is small for the relativistic solar axions considered here, the conversion remains coherent over the atomic scale and produces an electromagnetic signal whose deposited energy is equal to the incident axion energy.

The differential inverse Primakoff cross section is given by Refs.~\cite{Dent2020,Abe2021} to be:
\begin{equation}
\frac{d\sigma}{d\Omega}
=
\frac{\alpha g_{a\gamma}^{\,2}E_a^3p_a\sin^2\theta}
{4\pi\left(E_a^2+p_a^2-2E_ap_a\cos\theta\right)^2}
\,|Z-F(q)|^2,
\label{eq:primakoff_diff}
\end{equation}
where $E_a$ and $p_a$ are the axion energy and momentum, $\theta$ is the scattering angle, $Z$ is the atomic number, $F(q)$ is the atomic form factor, and $q$ is the momentum transfer. In the relativistic limit ($m_a\ll E_a$), appropriate for the solar axions studied here, Eq.~(\ref{eq:primakoff_diff}) simplifies to
\begin{equation}
\frac{d\sigma}{d\Omega}
=
\frac{\alpha g_{a\gamma}^{\,2}}
{16\pi}
\frac{\sin^2\theta}
{(1-\cos\theta)^2}
\,|Z-F(q)|^2.
\label{eq:primakoff_rel}
\end{equation}

The expected rate from solar axions is obtained by folding the transverse-plasmon axion flux of Eq.~(\ref{eq:TP_flux}) with the inverse Primakoff cross section:%, summing the contributions from sodium and iodine nuclei, and propagating the resulting spectrum through the detector response, including the energy resolution and detection efficiency. 
\begin{equation}\label{eqn:rate}
    R = \frac{d\Phi_a}{dE_a}\times
    \left(
    \frac{d\sigma}{d\Omega}_{\rm Na}N_{\rm Na}+\frac{d\sigma}{d\Omega}_{\rm I}N_{\rm I}
    \right)\times MT\times F_{det}(E),
\end{equation}
where $N_{\rm Na}, N_{\rm I}$ are the number of Na and I nuclei in the detector, $MT$ is the total detector mass and livetime exposure, and $F_{det}(E)$ is its response function accounting for both the energy resolution and efficiency.

Since the solar axion flux scales as the inverse square of the Earth--Sun distance,  a small annual modulation in the expected event rate is induced by the eccentricity \mbox{($\epsilon=0.0167$)} of the Earth's orbit ($T=1$ sidereal year period), reaching its maximum at perihelion \mbox{($t_0=$~January~3)}. As $\epsilon\cos(\frac{2\pi t-t_0}{T}) << 1$, we may expand the inverse-square dependence to second order and model the time-dependent part of the expected signal as
\begin{equation}
S(t)=
A\left[
2\epsilon\cos\!\left(\omega(t-t_0)\right)
+3\epsilon^2
\cos^2\!\left(\omega(t-t_0)\right)
\right],
\label{eq:modulation}
\end{equation}
where $\omega=2\pi/T$ and $A$ is the modulation amplitude.
For an average event rate $R_a$ evaluated at the mean Earth--Sun distance, the corresponding modulation amplitude is:
\begin{equation}
A=
\frac{R_a}{2}
\left[
\frac{1}{(1-\epsilon)^2}
-
\frac{1}{(1+\epsilon)^2}
\right].
\label{eq:amplitude}
\end{equation}

%These measured amplitudes are subsequently compared jjjjjjjjjjjjjjjjjjjjkkkkkjjj:q
%with the predicted amplitudes .

\begin{figure}%[!hbt]
    \begin{center}
    \includegraphics[width=1\columnwidth]{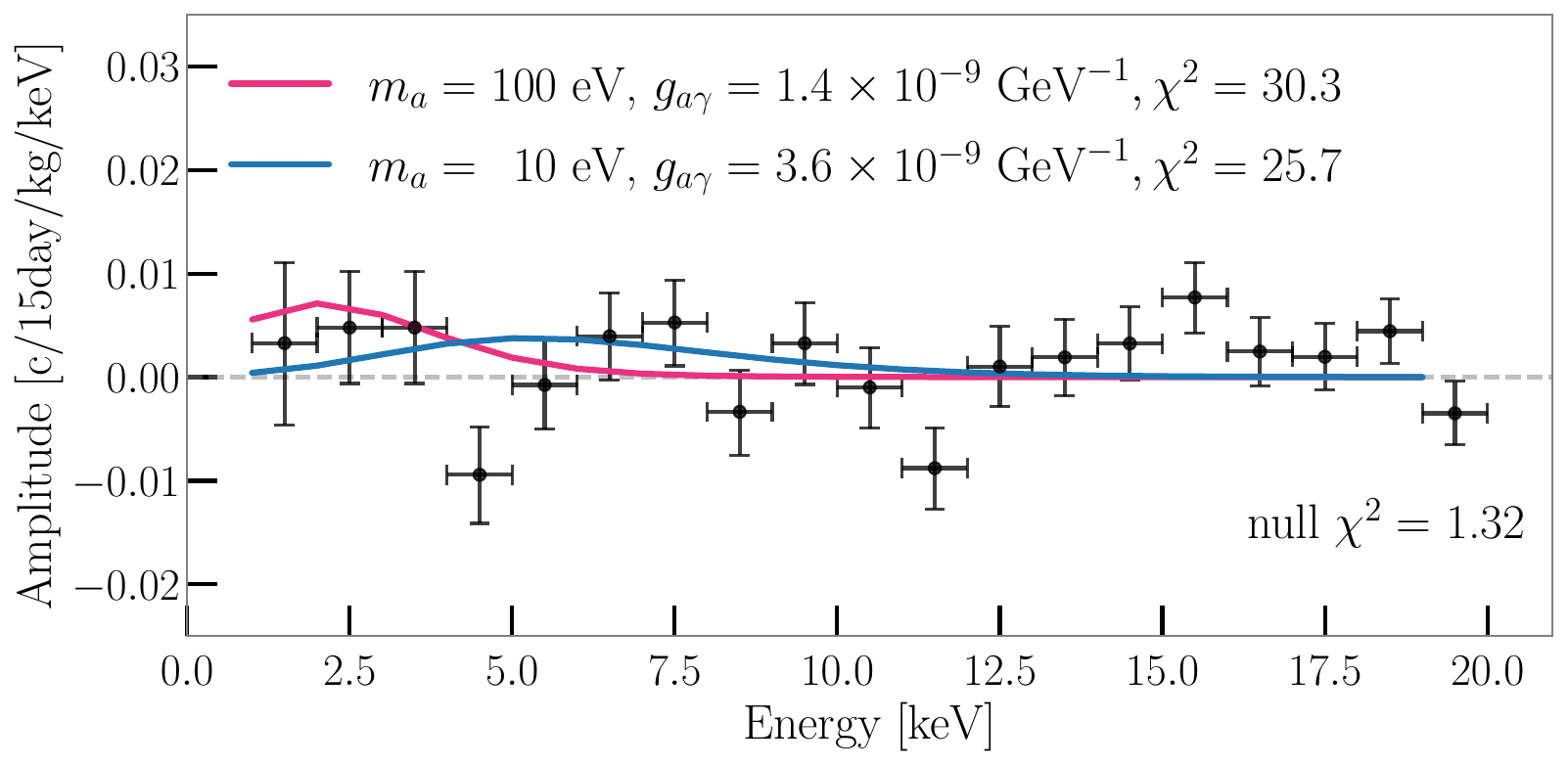}	 
    \caption{
Measured annual modulation amplitudes obtained from simultaneous fits in 1\,keV energy intervals. The grey dashed line corresponds to the null-modulation hypothesis. %All measured amplitudes are statistically consistent with zero, providing no evidence for the annual modulation expected from solar axions. 
The blue (magenta) curve shows the amplitudes expected of an axion with axion-photon coupling $g_{a\gamma}= 3.6\times10^{-9} (1.4\times10^{-9})$ \,GeV$^{-1}$ at axion mass $m_a=10(100)$\,eV where the Primakoff (TP) flux dominates the axion production.
%\mg{I just noticed a possible typo. Please, confirm. We write "at axion mass $m_a=100(10)$~eV where the Primakoff (transverse-plasmon) flux dominates". Perhaps, we mean the opposite. At 10 eV Primakoff dominates, while the TP dominates at 100 eV.}\SH{Yes that's right, I have changed it now}
}
	    \label{fig:modulations}
	\end{center}
\end{figure}

%As the Earth's orbit about the Sun is an off-center ellipse, the day corresponding to the shortest distance between the two bodies is January~3rd. The annual modulation in this distance would result in a time-dependent event rate in the experiment, tracing an expected signal of: 
%\begin{equation}
%    S(t_i) = A \left( \cos(\omega(t_i-t_0)) + \frac{3}{2}\epsilon\cos^2(\omega(t_i-t_0))\right),
%    \label{eqn:modulation}
%\end{equation}
%where $\omega = \frac{2\pi}{T}$ for the period of $T=$ 1~year, $t_0$ is the phase (fixed to January~3), and $A$ is the modulation amplitude. 

%The expected amplitude is derived from the difference in the rate ($R_a$ when the Earth is $1.496\times10^{11}$~m from the Sun) at aphelion and perihelion when the Earth is farthest and closest to the Sun, respectively. The amplitude of the modulation is written as:
%\begin{equation}
%    A = \frac{1}{2}\left(  \frac{R_a}{(1-\epsilon)^2} - \frac{R_a}{(1+\epsilon)^2} \right),
%    \label{eqn:modulation}
%\end{equation}
%where $\epsilon=0.0167$ is the orbit ellipticity. 

%%%%%%%%%%%%%%%%%%%%%%%%%%%%%%%%%%%%%%%%%%%
%%%%%%%%%%%%%%%%%%%%%%%%%%%%%%%%%%%%%%%%%%%
%\vspace{0.2cm}
%\textit{Detector background model and modulation search.---}
%\textit{Search strategy.---}
The expected annual modulation from solar axions is superimposed on a slowly varying detector background arising from the decay of radioactive isotopes from cosmogenic, crystal bulk, and detector material origins. An accurate description of this time-dependence is therefore essential to avoid biasing the modulation analysis. In ANAIS-112, the background evolution is modeled using a Geant4 simulation constrained by measured radioactive contamination of the NaI(Tl) crystals and detector components~\cite{ANAISPerformance,ANAIS_bkgd2021}. Further details on the ANAIS-112 background model are included in the Supplemental Material. 

The modulation search is performed independently in \SI{1}{keV}
energy bins between 1 and 20~keV. For each energy bin, the data from each detector are grouped into 15-day time bins. The
corresponding event rates for each detector are corrected by the event selection efficiency and normalized to the
live-time exposure in each bin.
For each detector $d$ and time bin $t_i$, the expected background is parameterized as
\begin{equation}
B_d(t_i)=
R_{0,d}
\left[
f_d\,\phi^{\rm MC}_{\rm bkg,d}(t_i)
+
(1-f_d)\,\phi_{\rm flat}(t_i)
\right],
\label{eqn:bkgd}
\end{equation}
where $\phi^{\rm MC}_{\rm bkg,d}$ describes the time evolution predicted by the Geant4 background model, while $\phi_{\rm flat}$ accounts for residual time-independent contributions, including a low-energy 
%noise 
component below \SI{3}{keV} which is unexplained by the background model (see Supplemental Material) and the average contribution of a possible ALP signal. The nuisance parameters $R_{0,d}$ and $f_d$ determine the overall normalization and the relative contribution of the simulated background evolution, respectively. This parameterization provides an excellent description of the long-term detector rates throughout the full data-taking period~\cite{ANAISPerformance,ANAIS_bkgd2021}.

The modulation amplitude is extracted independently in 1\,keV energy intervals by simultaneous fits of the sum of Equations~\eqref{eq:modulation} and~\eqref{eqn:bkgd} to the time evolution of the event rates in all detector modules. Unaccounted time variations in the detector efficiency or inaccuracies in the modeled background evolution could bias the inferred annual-modulation amplitude. The stability studies and systematic checks presented in~\cite{ANAIS6y, Amare:2021yyu} indicate that these effects are subdominant compared with the statistical uncertainty for the variations observed in ANAIS-112.
Figure~\ref{fig:modulations} shows the resulting modulation amplitudes between 1--20\,keV. Within the experimental uncertainties, all measured amplitudes are consistent with the null-modulation hypothesis, and no statistically significant annual modulation is observed.
Bayesian statistics are used to compare the expected modulation amplitudes with the measured values to derive constraints on the axion--photon coupling $g_{a\gamma}$.

For each axion mass and coupling hypothesis, we construct a Gaussian likelihood from the measured modulation amplitudes for 1--20\,keV.
%
%\begin{equation}
%\mathrm{LLH}=-\frac{1}{2}\sum_i\left[\frac{\left(A_i^{\rm obs}-A_i^{\rm pred}\right)^2}{\sigma_i^2}+\ln\!\left(2\pi\sigma_i^2\right)\right],
%\label{eq:LLH}
%\end{equation}
%
%where $A_i^{\rm obs}$ is the measured modulation amplitude in energy bin $i$, $A_i^{\rm pred}$ is the corresponding prediction for a given choice of axion couplings, and $\sigma_i$ is the associated statistical uncertainty.
The posterior probability distribution is obtained by exponentiating the likelihood and assuming flat priors on $g_{a\gamma}^4$, as the signal strength is proportional to $g_{a\gamma}^4$.
%\ms{priors were flat on g or in g4?}\SH{Priors are flat in g^4, following $\int L(g^4)dg^4$ }
%
%
%\begin{equation}
%P(\theta)=\frac{\mathcal{L}(\theta)}{\displaystyle\int \mathcal{L}(\theta')\,d\theta'},
%\label{eq:posterior}
%\end{equation}
%
%where $\theta$ denotes the parameter of interest. The 90\% credibility interval is defined by
%
%\begin{equation}
%\int_{0}^{\theta_{90}}P(\theta)\,d\theta=0.9.
%\label{eq:CL}
%\end{equation}
%\MM{All this is pretty standard. I wonder whether we need to include all these formulas. In any case, I'd remove the second term in Eq. (12), since it is just a normalization constant.}\SH{Yes I agree, I think we can just say we use the statistics without explaining in detail}
The strongest experimental upper limit corresponds to the high mass regime, where the transverse plasmon flux is dominant, for an axion mass of  $m_a\approx150$\,eV,
\begin{equation}
g_{a\gamma}<1.32\times10^{-9}\ {\rm GeV}^{-1}\qquad(90\%~{\rm C.L.}),
\label{eq:gag_limit}
\end{equation}
%\JRA{we should frame \ref{eq:gag_limit} in a given mass range. At higher energies we are better!!} \SH{I re-formulated the presentation of the result to give the value for the peak of our dip, maybe you can review the wording to see that it makes sense?}
%
while the complete exclusion limits as a function of the axion mass are presented in Figure~\ref{fig:exclusion}. In the resonant mass range of $35$--$280$\,eV, resonant transverse-plasmon production enhances the sensitivity to the axion--photon coupling beyond that achievable with the conventional Primakoff flux. The dominant systematic uncertainty, associated with the solar magnetic-field model, affects the derived exclusion limits by less than $20\%$. Experimental systematics associated with the detector response and background model are subdominant with respect to the theoretical uncertainty from the solar magnetic-field model. This work constitutes the first experimental exploration of this production mechanism.

%%%%%%%%%%%%%%%%%%%%%%%%%%%%%%%%%%%%%%%%%%%%%%%%%%%%%%%%%%%%%%%%%%%%%%%%%%%%%%
%\vspace{0.2cm}
\textit{Conclusions.---}
We have presented the first experimental search for solar axion-like particles produced through resonant transverse-plasmon conversion in the magnetic fields of the solar interior. Using 6.4-year exposure of the ANAIS-112 experiment, corresponding to $625.75\,\rm{kg\times yr}$, we searched for the characteristic annual modulation expected from this previously unexplored solar axion flux. % through both the inverse Primakoff and axio-electric detection channels. No statistically significant excess over the expected background was observed. 
No modulation inconsistent with the null-hypothesis was observed.

The absence of a signal allows us to derive the first experimental constraints on the axion--photon coupling associated with transverse-plasmon production in the Sun, with the strongest limit, $g_{a\gamma}<1.32\times10^{-9}\,\mathrm{GeV}^{-1}$ at 90\%~C.L., reached at $m_a\simeq150$\,eV.
%\mg{I added just a small clarification here, with the resutls. For those who read just the conclusions}. 
The characteristic shape of the exclusion curve is a direct consequence of the resonant nature of the production mechanism, with the interval 35--280~eV representing the mass range over which the TP flux exceeds the Primakoff flux for the solar magnetic-field model adopted here.
%For the inverse Primakoff channel, ANAIS-112 establishes the most stringent laboratory limits in the axion mass range between approximately 80 and 220~eV, improving upon the previous best experimental constraints by up to $22\%$, corresponding to a factor of 1.28 in the excluded coupling. %We also present complementary limits in the $g_{ae}g_{a\gamma}$ parameter space from the axio-electric detection channel.

These results establish resonant transverse-plasmon conversion as a viable and experimentally testable source of solar axions, expanding the range of production mechanisms accessible to terrestrial searches. %More generally, they demonstrate that low-background rare-event experiments, originally developed for dark matter detection, can provide competitive sensitivity to previously unexplored solar axion production channels. 
Moreover, consideration of the TP flux may re-evaluate the interpretation of the axion rates expected as illustrated in Figure~\ref{fig:modulations} where a signal from TP flux dominated mass regime would peak at roughly 2\,keV compared to the Primakoff $\approx6$\,keV peak.

This Letter opens the experimental exploration of resonant transverse-plasmon production as a source of solar axions and demonstrates that existing low-background underground experiments can already probe this production mechanism, establishing a new framework for future helioscopes and low-background dark matter experiments.

%Suggested text from Jaime
%This Letter opens the experimental exploration of resonant transverse-plasmon production as a source of solar axions and demonstrates that existing low-background underground experiments can already probe this production mechanism, establishing a new framework for future helioscopes and low-background dark matter experiments.

The data underlying the plots and other findings presented in this study are available upon request to the authors~\cite{rama_git, DMDC}.

%%%%%%%%%%%%%%%%%%%%%%%%%%%%%%%%%%%%%%%%%%%%%%%%%%%%%%%%%%%%%%%%%%%%%%%%%%%%%%
\vspace{0.2cm}

\textit{Acknowledgments.---}
%\jra{ANAIS, ZGZ, etc ... include project numbers, funding, etc.}
The ANAIS-112 experiment has been financially supported by MCIN/AEI/10.13039/501100011033 under
grants PID2022-138357NB-C21, PID2019-104374GB-I00, and FPA2017-83133-P as well as by the Consolider-Ingenio 2010 Programme through the MultiDark (CSD2009-00064) project. Additional support was provided by the LSC Consortium, the Gobierno de Aragón,
and the European Social Fund (Group in Nuclear and Astroparticle Physics), as well as funds from the European Union through the NextGenerationEU/PRTR initiative (Planes Complementarios, Programa de Astrofísica y Física de Altas Energías). The ANAIS authors gratefully
acknowledge the use of the Servicio General de Apoyo a la Investigación – SAI of the Universidad de Zaragoza, as well as the technical support provided by the staff of LSC and GIFNA.
J.\,Ruz and J.\,K.\,Vogel acknowledge support from the 
Deutsche Forschungsgemeinschaft (DFG, German Research Foundation) under Germany’s Excellence Strategy – Cluster of Excellence “Color meets Flavor”, EXC 3107 – Project-ID 533766364. J.\,K.\,Vogel also acknowledges funding from the German federal and state program "Professorinnenprogramm 2030" Project-ID 01FP24167Q. R.\,M.\,Alkaddah acknowledges support from Erasmus Mundus Joint Master (EMJM) Scholarship from the European Commission. 
M.\,Giannotti acknowledges support from the Spanish Agencia Estatal de Investigación through grant PID2019-108122GB-C31, funded by MCIN/AEI/10.13039/501100011033, and from the European Union NextGenerationEU/PRTR initiative (Planes Complementarios, Programa de Astrofísica y Física de Altas Energías). M.\,Giannotti also acknowledges support from grant PGC2022-126078NB-C21, `Aún más allá de los modelos estándar,'' funded by MCIN/AEI/10.13039/501100011033 and by the European Regional Development Fund (ERDF), `A way of making Europe''; from the European Union's Horizon 2020 research and innovation programme under European Research Council (ERC) grant agreement No.\ 788781 (IAXO+); and from project CNS2025-165965, ``Señales de axiones en el rango MeV desde fuentes astrofísicas hasta detectores actuales y futuros,'' funded by MICIU/AEI/10.13039/501100011033.

%%%%%%%%%%%%%%%%%%%%%%%%%%%%%%%%%%%%%%%%%%
%%%%%%%%%%%%%%%%%%%%%%%%%%%%%%%%%%%%%%%%%%
%%%%%%%%%%%%%%%%%%%%%%%%%%%%%%%%%%%%%%%%%%
%%%%%%%%%%%%%%%%%%%%%%%%%%%%%%%%%%%%%%%%%
%%%%%%%%%%%%%%%%%%%%%%%%%%%%%%%%%%%%%%%%
\bibliographystyle{apsrev4-2}
\bibliography{mybib}

%%%%%%%%%%%%%%%%%%%%%%%%%%%%%%%%%%%%%
%%%%%%%%%%%%%%%%%%%%%%%%%%%%%%%%%%%%
\clearpage     
%%%%%%%%%%%%%%%%%%%%%%%%%%%%%%%%%%%%%%%%%%%%%%%%%%%%%%%%%%%%%%%%%%%%%%%%%%%%%%
%% Supplemental Material
%%%%%%%%%%%%%%%%%%%%%%%%%%%%%%%%%%%%%%%%%%%%%%%%%%%%%%%%%%%%%%%%%%%%%%%%%%%%%%

\setcounter{page}{1}
\setcounter{figure}{0}
\setcounter{equation}{0}
\setcounter{table}{0}

\renewcommand{\theequation}{S\arabic{equation}}
\renewcommand{\thepage}{S\arabic{page}}
\renewcommand{\thefigure}{\Alph{figure}}

\clearpage
\onecolumngrid

\begin{center}
    \Large\textbf{Supplemental Material}
\end{center}

\vspace{0.5cm}

\twocolumngrid

\title{Supplemental Material:\\
Transverse Plasmon--Axion Production in the Solar Interior}

\maketitle

This Supplemental Material provides the theoretical details supporting the analysis presented in the main Letter. We summarize the solar magnetic-field models adopted throughout this work, review the plasma dispersion relations governing plasmon--axion mixing and derive the resonant transverse-plasmon axion flux employed in the analysis. %, and discuss the treatment of off-resonant production. 
We also describe the ANAIS-112 time-dependent background model and illustrate the simultaneous background-plus-modulation fit for a representative energy interval. Unless otherwise stated, the notation follows that of the main text.

%%%%%%%%%%%%%%%%%%%%%%%%%%%%%%%%%%%%%%%%%%%%%%%%%%%%%%%%%%%%%%%%%%%%%%%%%%%%%%

%%%%%%%%%%%%%%%%%%%%%%%%%%%%%%%%%%%%%%%%%%%%%%%%%%%%%%%%%%%%%%%%%%%%%%%%%%%%%%
\section{Solar magnetic-field models}\label{sec:Bfield}

The resonant TP axion flux is computed using solar magnetic-field models~\cite{Guarini2020}, as implemented in the open-source \texttt{SolarAxionFlux} framework~\cite{Hoof}. The magnetic field of the solar interior cannot be measured directly, and helioseismology, together with the observed solar oblateness, provides upper bounds rather than a determination~\cite{Antia:2000pu,Friedland:2002is,Couvidat:2002gvk}. The benchmark field profiles adopted here are chosen to saturate these observational limits and should not be interpreted as measurements of the internal solar magnetic field.
%\mg{This is my main comment on the section. ``Helioseismologically constrained'' seems to indicate that the field is known. the radiative-zone bound quoted in the literature runs from $\sim$5--7~MG (Friedland \& Gruzinov, from solar oblateness) to $\sim$30~MG (Couvidat et al.).}.

\begin{figure}[!b]
    \centering
    \includegraphics[width=0.95\linewidth]{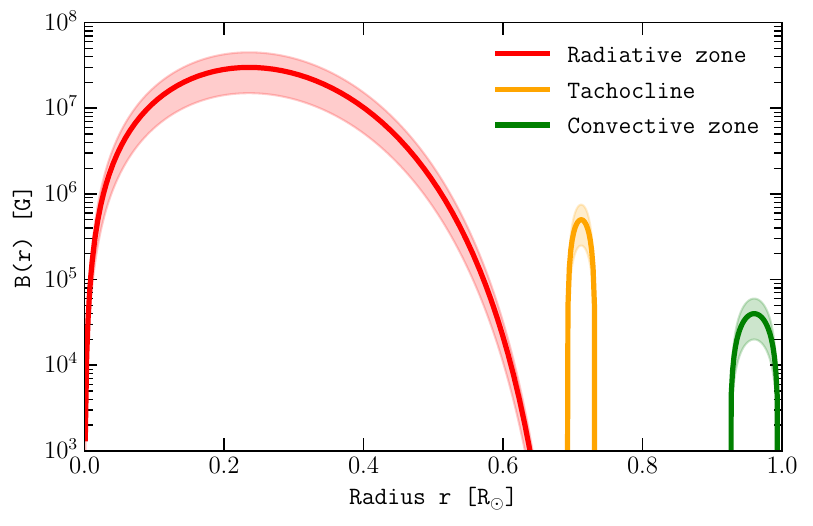}
    \caption{
   Radial profile of the seismic solar magnetic-field model adopted in this work~\cite{Guarini2020,Hoof}. The red, yellow, and green regions correspond to the radiative zone, tachocline, and convective zone, respectively. The shaded band represents the magnetic-field uncertainty propagated to the predicted TP axion flux.
    }
    \label{fig:solar_B_profile}
\end{figure}

These models span the range of solar magnetic-field strengths compatible with helioseismological constraints. Throughout this work, we adopt the seismic model, in which the magnetic field is decomposed into contributions from the radiative zone, tachocline, and convective envelope, with peak strengths of approximately [3$\times10^{7}$], [5$\times10^{5}$] and [5$\times10^{4}$]~G, respectively. The corresponding radial profile and propagated uncertainty band are shown in Figure~\ref{fig:solar_B_profile}. The magnetic-field strength at the resonance radius then sets the normalization of the resonant axion flux. 
The mass range of interest here, \mbox{$m_a\simeq35$--$280$\,eV}, resonates in the radiative zone, where the field is of fossil origin and is not expected to vary on solar-cycle timescales. The convective-envelope field, which does vary over the eleven-year cycle, contributes to the resonant flux only for $m_a\lesssim$ few~eV, where the transverse-plasmon flux is in any case negligible compared with the Primakoff flux (Figure~\ref{fig:TP_fluxes}). The predicted signal is therefore static over the ANAIS-112 data-taking period, and the only time dependence is the annual one induced by the Earth--Sun distance.

%%%%%%%%%%%%%%%%%%%%%%%%%%%%%%%%%%%%%%%%%%%%%%%%%%%%%%%%%%%%%%%%%%%%%%%%%%%%%%
\section{Transverse-plasmon axion production and flux on Earth}

Resonant plasmon--axion production occurs when the local solar plasma frequency equals the axion mass, Eq.~(\ref{eq:omegap_res}), selecting a narrow radial shell within the solar interior~\cite{Guarini2020}. Consequently, different axion masses probe different regions of the Sun, each characterized by distinct magnetic-field strengths and plasma conditions. This resonance structure determines both the normalization and the spectral shape of the emitted axion flux.

\begin{figure}[!b]
    \centering
    \includegraphics[width=\linewidth]{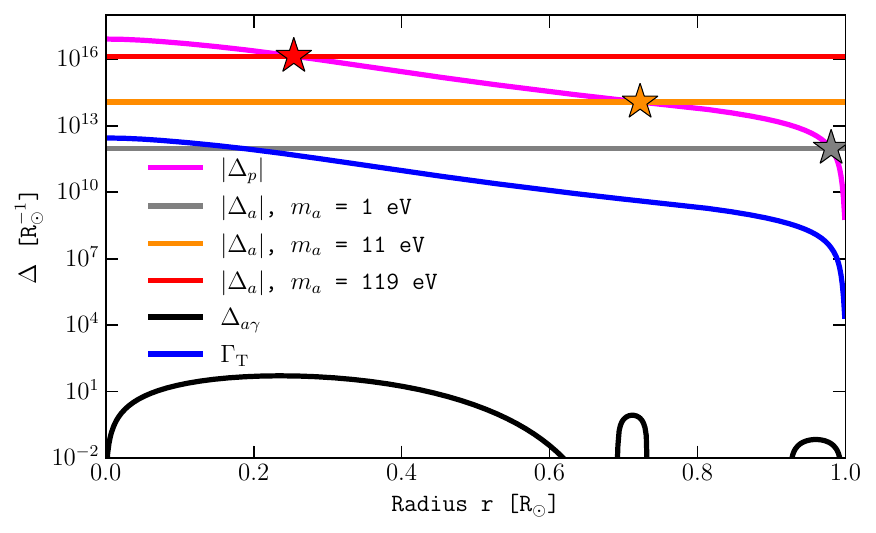}
    \caption{
Radial dependence of the photon--axion oscillation parameters entering the resonant conversion probability. The intersections of $|\Delta_p|$ and $|\Delta_a|$ (stars) satisfy the resonance condition, Eq.~\eqref{eq:omegap_res}, defining the resonant shells for $m_a=1$, $11$, and $119\,\mathrm{eV}$. Also shown are the photon--axion mixing term $\Delta_{a\gamma}^{T}$ and the transverse photon the damping rate $\Gamma_T$.
}
    \label{fig:axion_terms}
\end{figure}

Figure~\ref{fig:axion_terms} illustrates the radial dependence of the oscillation parameters introduced in Eq.~(\ref{eq:deltas}) for representative axion masses. The plasma contribution, $|\Delta_p|$, decreases monotonically from the solar core towards the surface following the electron-density profile, whereas the axion term, $|\Delta_a|$, remains constant throughout the Sun for a fixed axion mass. Their intersections (star symbols), corresponding to the resonance condition $\Delta_p=\Delta_a$, identify the resonant conversion shells where photon--axion mixing is maximized. 

As expected from Eq.~(\ref{eq:deltas}), increasing the axion mass shifts the resonance to regions of higher plasma frequency and therefore deeper into the solar interior. Figure~\ref{fig:axion_terms} also displays the photon--axion mixing term,
\begin{equation}
     \Delta_{a\gamma}^T = \frac{g_{a\gamma}B_T}{2},
\end{equation}
and the photon absorption rate, since thermal
photons are continuously emitted and reabsorbed in the
Sun. This process is characterized by the transverse photon damping rate $\Gamma_T$ \cite{RedondoRaffelt}.
Throughout the solar interior,
$\Delta_{a\gamma}\ll |\Delta_p|$,
$\Delta_{a\gamma}\ll |\Delta_a|$,
and
$\Delta_{a\gamma}\ll\Gamma_{\rm abs}$.
The system therefore remains in the weak-mixing regime, while the finite photon absorption width regularizes the resonance and sets its effective width.

%%%%%%%%%%%%%%%%%%%%%%%%%%%%%%%%%%%%%%%%%%%%%%%%%%%%%%%%%%%%%%%%%%%%%%%%%%%%%%
%\section{Transverse-plasmon axion flux}

Eq.~\eqref{eq:TP_flux} is evaluated by integrating the resonant production rate over the solar volume to obtain the differential axion flux at Earth. Because the resonance condition in Eq.~\eqref{eq:omegap_res} is satisfied only within a narrow spherical shell, the integration is restricted to this region using the Gaussian window
\begin{equation}
\label{eq:gaussian_window}
W(r)
=
\exp\!\left[
-\frac{\left|\omega_p(r)-m_a\right|^2}{W_0^2}
\right],
\end{equation}
where $W_0$ is chosen proportional to the local resonance width. The calculations are performed with the open-source \texttt{SolarAxionFlux} package~\cite{Hoof}, extended to identify the resonance shell automatically for each axion mass while retaining the finite-width treatment of resonant transverse-plasmon production~\cite{rama_git}.

\begin{figure}[!h]
    \centering
    \includegraphics[width=\linewidth]{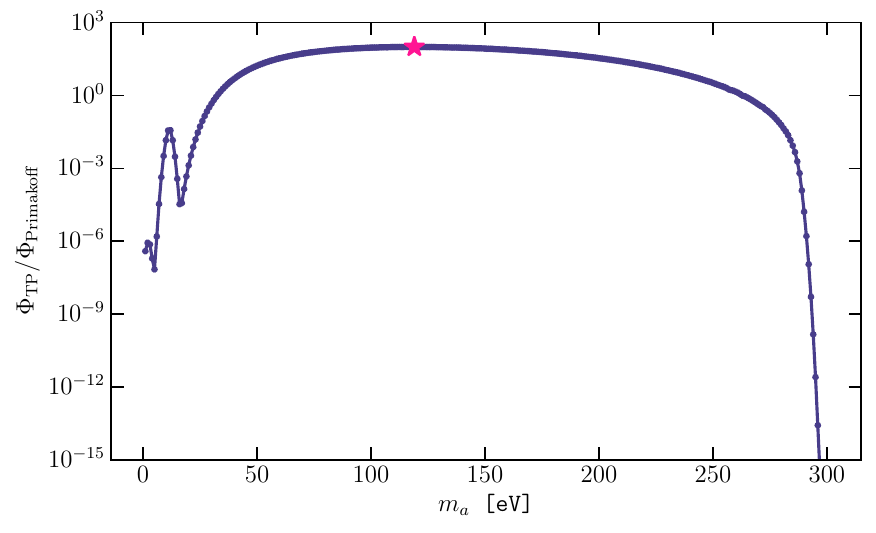}
    \caption{
  Ratio of the resonant transverse-plasmon to Primakoff axion flux as a function of the axion mass. The maximum ratio is reached at $m_a=119~\mathrm{eV}$, indicated by the star.
    }
    \label{fig:TP_fluxes}
\end{figure}

Figure~\ref{fig:TP_fluxes} shows the ratio of the TP to Primakoff axion fluxes as a function of the axion mass. The resonant contribution reaches enhancements of up to two orders of magnitude relative to the Primakoff flux.

As discussed in the main text, the dominant systematic uncertainty arises from the adopted $\pm50\%$ uncertainty in the solar magnetic field~\cite{Guarini2020,Hoof}. Its impact on the inferred axion--photon coupling is evaluated by repeating the complete analysis for the different magnetic-field configurations described above. The resulting variation is summarized in Table~\ref{tab:systematic}, yielding an asymmetric systematic uncertainty of ${}^{+19\%}_{-10\%}$ on the excluded axion--photon coupling.

\begin{figure}[!h]
    \centering
    \includegraphics[width=1.0\linewidth]{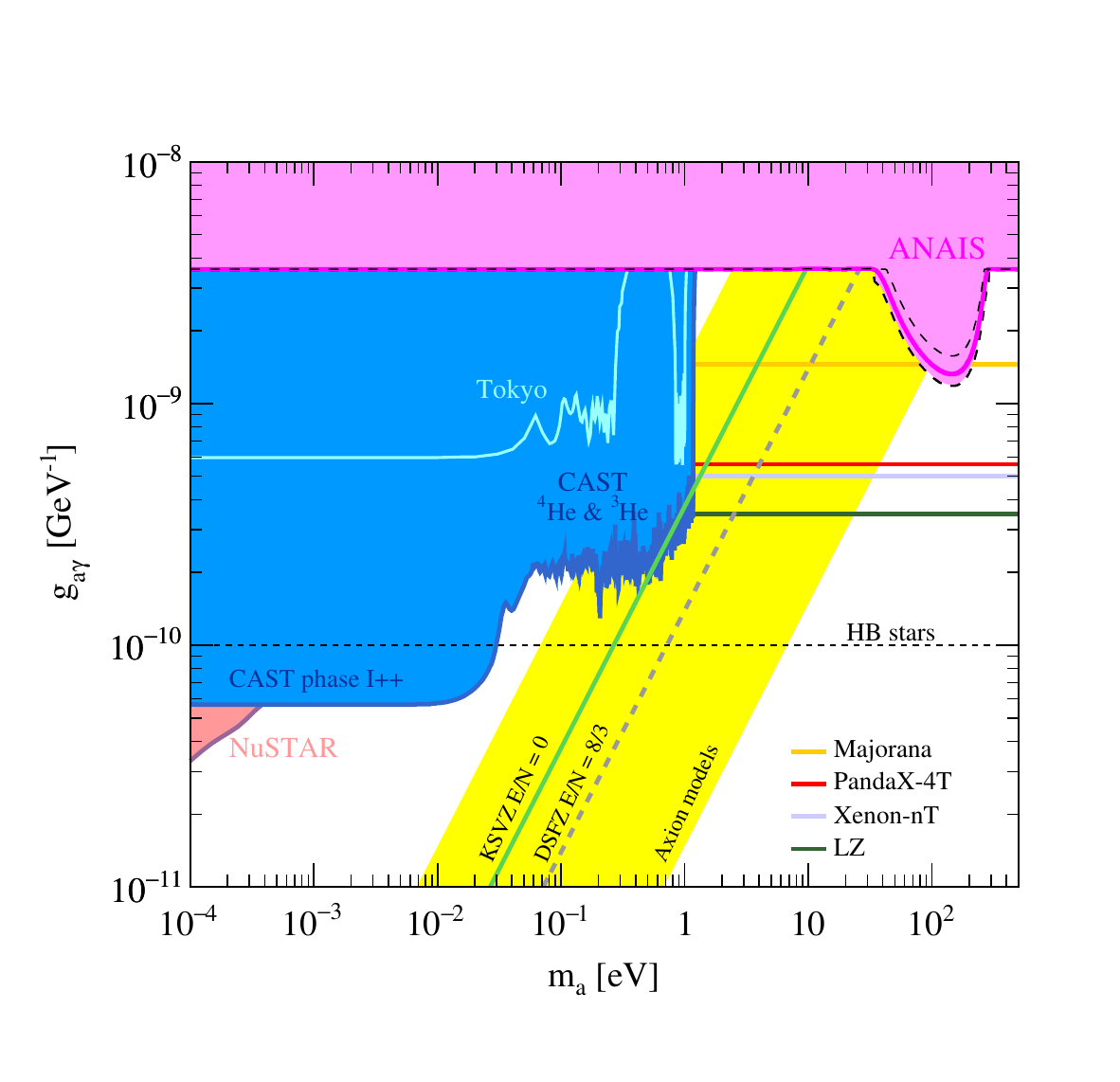}
    \caption{
Expanded view of the exclusion limits, highlighting the ANAIS-112 result in comparison with CAST, NuSTAR, and other leading experimental searches. The dotted lines show the exclusion band obtained by propagating the uncertainty in the solar magnetic-field model. Representative QCD axion models are also displayed.
}
    \label{fig:exclusion_sys}
\end{figure}

Figure~\ref{fig:exclusion_sys} illustrates this effect, where the dashed band represents the envelope of exclusion limits obtained for the different magnetic-field models. The figure demonstrates that, although the uncertainty in the solar magnetic field constitutes the dominant theoretical systematic in the flux prediction, its impact on the final TPC exclusion limit remains below the $\mathcal{O}(20\%)$ level. 

\begin{table}[!h]
    \centering
    \begin{tabular}{lc}
         \toprule
         \textbf{Source} & \textbf{Effect on $g_{a\gamma}$} \\
         \midrule
         Solar magnetic field & $^{+19\%}_{-10\%}$ \\
         \bottomrule
    \end{tabular}
    \caption{Dominant systematic uncertainty on the inferred axion--photon coupling for $m_a=0$, arising from the choice of solar magnetic-field model.}
    \label{tab:systematic}
\end{table}

%%%%%%%%%%%%%%%%%%
%%%%%%%%%%%%%%%%%%%
%\newpage

\section{ANAIS-112 time-dependent Background Model}\label{supp_anaisbkgd}

%\MM{I've enlarged this section including also information of the bkg model construction and the low-energy noise events that are mentioned in the main text}
The ANAIS-112 event rate exhibits a slowly varying, time-dependent background dominated by radioactive
contaminants in the detector. In the 1--20\,keV energy region considered in this work, the main contributions
originate from radioactive isotopes in the NaI(Tl) crystal bulk, cosmogenically activated isotopes, the photomultiplier
tubes (PMTs), and other detector components. The most
relevant contributions include $^{210}$Pb ($T_{1/2}=22.3$\,yr), both on the 
crystal surfaces and in the bulk, $^{40}$K ($T_{1/2}=1.25\times10^{9}$\,yr), and cosmogenically produced $^{22}$Na ($T_{1/2}=2.60$\,yr) and $^{3}$H ($T_{1/2}=12.32$\,yr). Shorter-lived cosmogenic
isotopes of iodine and tellurium are also relevant,
particularly during the first years of operation of the most
recently installed modules. Accurately accounting for this
time evolution is essential to avoid biasing the annual
modulation search~\cite{Amare:2021yyu, Hafizh_paper}.

The ANAIS-112 background model is based on a detailed
\textsc{Geant4} description of the nine detector modules,
their copper encapsulation, quartz optical windows,
photomultipliers, shielding, and surrounding components~\cite{ANAISPerformance}. An improved description of the ANAIS-112 background model is provided in~\cite{tesisTamara}.
The activities of the relevant radioactive isotopes are
independently constrained using several experimental
techniques. 
Internal $^{40}$K, $^{22}$Na, and other cosmogenic isotopes are quantified through coincidence measurements between detector modules. The $^{210}$Pb, $^{238}$U, and $^{232}$Th activities are constrained using alpha rates and Bi--Po coincidences.
%The $^{210}$Pb activity is inferred primarily from the measured alpha rate and the evolution of the $^{210}$Pb--$^{210}$Bi--$^{210}$Po sequence, while the activities of the $^{238}$U and $^{232}$Th chains are constrained using Bi--Po coincidences. 
Radioactivity in
external detector components is determined by high-purity
germanium spectroscopy, with the PMTs providing the most
relevant external contribution at low energy. These
independently measured activities are used as inputs to the
simulation rather than being freely adjusted in the
modulation fit~\cite{ANAISPerformance}.

For each detector and energy interval, the simulation
provides the expected energy spectrum and temporal
evolution of the radioactive background. Long-lived
isotopes such as $^{210}$Pb and $^{3}$H produce a slow
decrease over the data-taking period, whereas $^{22}$Na
and the shorter-lived cosmogenic isotopes decrease more
rapidly. The simulated energy depositions are propagated
through the detector response, including the energy
resolution, to construct the
detector-dependent temporal templates
$\phi^{\mathrm{MC}}_{\mathrm{bkg},d}(t)$ used in Eq.~\eqref{eqn:bkgd} of
the main Letter. 
%The nuisance parameter $f_d$ determines the relative weight of the time evolution predicted by the simulation with respect to the time-independent component, while $R_{0,d}$ sets the overall event rate in each detector. This treatment preserves the independently predicted temporal evolution of the radioactive background while accommodating residual contributions not included in the simulation.

In addition to radioactive backgrounds, the ANAIS-112
data contain populations of low-energy light events that
do not originate from scintillation in the NaI(Tl) crystal
bulk. These include a readily identifiable population
produced mainly by Cherenkov emission in the PMTs, as
well as a more challenging population of anomalous events
that frequently exhibits an asymmetric sharing of light
between the two photomultipliers. Both populations are
suppressed using the boosted-decision-tree (BDT) filtering
procedure developed for the ANAIS-112 annual modulation
analysis~\cite{Coarasa:2022zak,Coarasa:2024xec}.

%The BDT classifier is trained using $^{252}$Cf calibration events in the 1--2~keV interval, which predominantly correspond to bulk nuclear-recoil scintillation, as the signal sample. Non-NaI(Tl) light events recorded with the blank module in an equivalent energy interval are used as the noise sample. Events from the physics data set are not used in the training. The acceptance efficiency of the filter is determined independently from $^{252}$Cf and $^{109}$Cd calibration data and monitored throughout the data-taking period. This procedure efficiently rejects non-NaI(Tl) light events while retaining a high acceptance for genuine bulk scintillation events.

After application of the BDT filter, the measured event
rate below 3\,keV still exceeds the prediction of the
radioactive-background model. The origin of this residual
low-energy contribution is under investigation. No
significant time dependence has been identified for this
component. It is therefore represented by the
time-independent term $\phi_{\mathrm{flat}}(t)$ in Eq.~\eqref{eqn:bkgd},
together with the average contribution of a possible ALP
signal.

\begin{figure*}[!t]
    \begin{center}
    \includegraphics[width=0.8\textwidth]{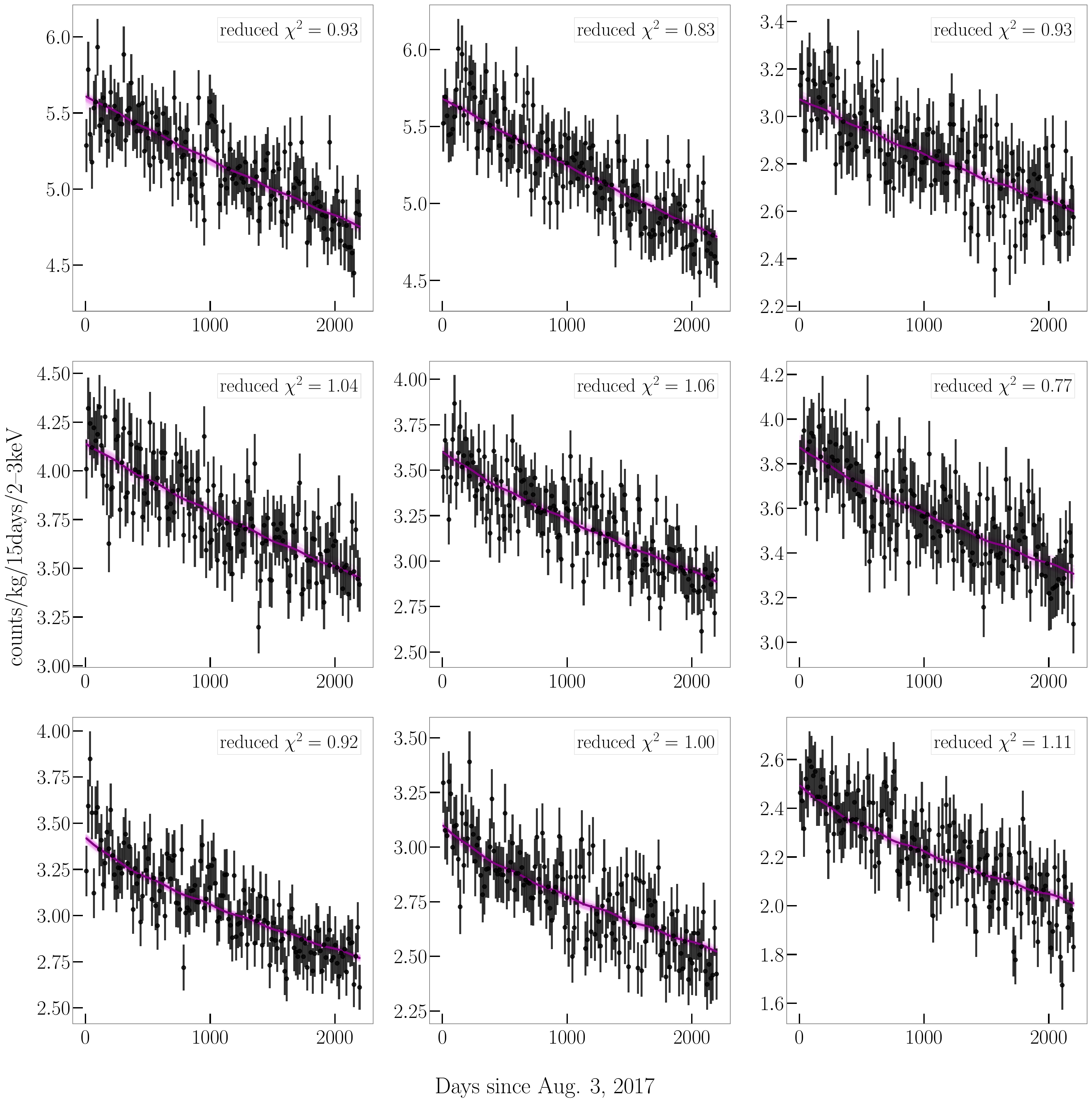}
	    \caption{The event rates (black points) measured in all nine ANAIS-112 crystals in the example energy region of 2--3~keV. The simultaneous fit (magenta curve) yields a modulation amplitude $A = (0.0047 \pm 0.0055)\, \mathrm{cpd\,kg^{-1}\,keV^{-1}}$ which is consistent with the no-modulation hypothesis. The reduced $\chi^2$ values shown in the individual panels indicate that the model provides an adequate description of the time evolution in all nine detectors.
        %\SH{Is it too big now?}
        }
	    \label{fig:rates}
	\end{center}
\end{figure*}

Figure~\ref{fig:rates} illustrates the
modulation analysis in a representative 
energy interval (2--3~keV). The event rates measured in the nine
ANAIS-112 modules are fitted simultaneously using the
background-plus-modulation model described in the main
Letter. The background normalization and the relative
contributions of the simulated and time-independent
components are allowed to vary independently for each
detector, while the modulation amplitude is common to all
modules. The modulation period and phase are fixed to one
year and the date of perihelion, respectively.

\end{document}